\documentstyle[osa,12pt]{revtex}

\begin{document}
\title{Test of a simple and flexible molecule model for $\alpha $-, $\beta $- and $%
\gamma $-S$_8$ crystals}
\author{C. Pastorino and Z. Gamba}
\address{Department of Physics, Comisi\'on Nacional de Energ\'\i a At\'omica, Av.\\
Libertador 8250, (1429) Buenos Aires, Argentina. e-mails:\\
clopasto@cnea.gov.ar 
and gamba@cnea.gov.ar\\}
\maketitle

\begin{abstract}

S$_8$ is the most stable compound of elemental sulfur in solid and liquid phases, at
ambient pressure and below 400K. Three crystalline phases of S$_8$ have been clearly
identified in this range of thermodynamic parameters, although no 
calculation of its phase diagram has been performed yet. 
$\alpha $- and $\gamma $-S$_8$ are orientationally ordered crystals while
$\beta $-S$_8$ is measured as orientationally disordered. In this 
paper we analyze the phase diagram of S$_8$ crystals, as given by
a simple and flexible molecule model, $ via $ a series of 
molecular dynamics (MD) simulations.
 The calculations are performed
in the constant pressure- constant temperature ensemble, using 
an algorithm that is able to reproduce structural phase transitions.\\ 
\end{abstract}

{\bf Introduction:}

Elemental sulfur shows a rich variety of molecular forms and therefore of
chemical and physical properties in the crystalline and liquid phases \cite
{steudel,meyer1}. Recently, a high pressure superconductor transition 
has been found, with T$_c$=17K (at 162GPa), the highest known for
any elemental solid \cite{highP-S}. Natural sulfur is of relevance in 
geophysics, astrophysics, material sciences and also of massive industrial 
use \cite{steudel}. Nevertheless, the complex phase diagram and properties
of its condensed phases are far from well known \cite{steudel,meyer1}.

The most stable
sulfur allotrope at ambient temperature and pressure (STP) is S$_8$, a
crown-shaped cyclic S$_8$ molecule, stable in solid, liquid and gas phases.

Orthorhombic $\alpha $-S$_8$ is the crystalline form stable at STP.
Figs 1a shows its "crankshaft" structure with four molecules in the 
primitive unit cell. The space group is D$_{2h}^{24}$ (Fddd), with 
16 molecules in this non-primitive orthorhombic cell. If 
$\alpha $-S$_8$ is slowly heated, it shows a phase transition to monoclinic 
$\beta $-S$_8$ at 369K (Fig. 1b), which melts at 393K \cite{steudel}. 
Nevertheless, 
most $\alpha $-S$_8$ crystals do no easily convert to $\beta $-S$_8$, 
they melt, instead, at
385.8K \cite{steudel,meyer1}. $\beta $-S$_8$ is a monoclinic crystal, with
six molecules in the primitive unit cell, two of them orientationally disordered
\cite{beta1,beta2}. A third crystalline allotrope has been observed: 
$\gamma $-S$_8$ (Fig. 1c), that can be obtained from solutions of S$_8$ or from its
melt \cite{steudel,meyer1}. The space group is C$_{2h}^{4}$ (P2/c), 
 with four molecules in the pseudo-hexagonal closed-packed unit
cell \cite{gam1}. The density of this allotrope, at STP, is 5.8 \%
higher than that of $\alpha $-S$_8$.

In spite of its relevance, calculations on the crystalline phase diagram of
elemental sulfur molecules are scarce. Previous molecular dynamics (MD) 
simulations of crystalline S$_8$ were limited to constant-volume simulations
at normal pressure and temperatures of $\alpha $-S$_8$ \cite{cardini1}, with
the only aim to study the intramolecular frequencies obtained by Raman and
ir. measurements. The same scope was followed by numerous works that measured
crystalline frequencies and performed calculations based on 
lattice dynamics sums in different aproximations \cite{alfafreq}. No attempt 
to reproduce the phase diagram of this
molecule, including determination of crystalline structures, lattice
dynamics, or thermodynamic properties of $\beta $- and $\gamma $-S$_8$ 
as a function of temperature
 was intended. Here we study the phase diagram of S$_8$ crystals, as given by
a simple model of a flexible molecule, by performing a series of constant pressure 
- constant temperature MD simulations, at several temperatures and zero pressure.\\

{\bf The inter- and intramolecular potential model:}

 The main problem in this type of calculation 
is the inter- and intramolecular potential model of this flexible molecule.
The above mentioned constant volume MD simulations of
 $\alpha $-S$_8$ were performed with an intermolecular potential model 
of the atom-atom type, to study the splitting of intramolecular 
frequencies in a sample of 64 molecules of $\alpha $ -S$_8$ \cite{cardini1}. 
 In our constant temperature - constant pressure MD simulations,
these atom-atom parameters are slightly changed to improve the fit of
our results to the
experimental unit cell volume and the estimated configurational energy of $%
\alpha $-S$_8$ at 300K and 0kbar pressure. The parameters of our LJ
potential model, for S-S interactions are: $\varepsilon $= 1.70kJ/mol and $%
\sigma $=3.39 \AA .

In a similar way, our intramolecular potential parameters
for bending and torsional angles of the flexible S$_8$ molecule are
based on those of ref. \cite{cardini1}. 
Our molecule model takes into account all intramolecular modes
that mix with lattice modes and can therefore be
relevant for structural phase transitions. As calculated in 
ref. \cite{cardini1} and also measured in all crystalline phases
at all temperatures, three main regions are
clearly separated in the vibrational spectra: between 10 
and $\sim $100cm$^{-1}$ are found the lattice and torsional 
modes, between 150 and 250cm$^{-1}$ the
bending modes and from 400 to 500cm$^{-1}$ the stretching modes, well
above in energies than the rest of vibrational modes. 
The bond length is therefore held constant at 2.060\AA\  in our
calculations.  Our flexible molecule model allows, instead,
 the mixing of the bending 
and torsional intramolecular motions with lattice modes in a straightforward
way. The intramolecular potential for bending angles S-S-S is taken harmonic

\[
V(\beta )=\frac 12C_\beta (\beta -\beta _0)^2, 
\]
with a force constant of C$_\beta $=25725k$_B/$rad$^2$ and $\beta _0$= 
108deg., k$_B$ is the Boltzmann constant. The intramolecular 
potential for torsion ($\tau $) angles is a double well:

\[
V(\tau )=A_\tau +B_\tau \cos (\tau )+C_\tau \cos ^2(\tau )+D_\tau \cos
^3(\tau )\text{,} 
\]
with $A_\tau $=57.192 k$_B$, $B_\tau $=738.415 k$_B$, $C_\tau $=2297.880 k$_B$ 
and $D_\tau $=557.255 k$_B$. These parameters describe a double well with minima at 
$\tau $=${^+_-}$ 98.8deg., and a barrier height of about 9kJ/mol at 
$\tau $=180deg. The barrier height at $\tau $=0deg. is of 30kJ/mol, out of
the range of energies explored in these simulations.\\

{\bf Calculations:}

The phase diagram and dynamical properties of S$_8$ crystals, as given
by this molecular model, are studied in the (N,P,T) ensemble,
 at zero pressure, by a series of classical 
 MD simulations. The external pressure
is considered isotropic and the MD algorithm allows
volume and shape fluctuations of the sample, in order to balance the
 internal stresses with an external isotropic pressure \cite{algor2}. The 
algorithm controls pressure $ via $ an 
 extended system which includes as
extra variables the MD box parameters. The temperature control of the sample
follows the approach of Nos\'e \cite{algor3,algor4,algor5}, which also 
includes an external variable.
The equations of motion of these flexible molecules are
integrated using the Verlet algorithm for the atomic displacements and the 
Shake algorithm for the constant bond length  
constraints on each molecule\cite{algor1,tildesley}.  The
final MD\ algorithm is totally similar to that used in a study of black
Newton films \cite{bubbles1}.

As the starting point of our simulations, we considered the available structural 
data at STP. The molecules were located at the experimental sites, with 
the  measured orientation but with their isolated D4d symmetry. In the
 course of
the simulation, the molecules distort to attain, on a time average, their
site symmetry. Following this procedure, the only problem found was with the 
$\beta $-S$_8$ sample. In this case the calculations were initiated 
with the LJ parameter $\varepsilon $ multiplied by a factor that in a few
(about 100) time steps was varied from 0.1 to 1. This procedure 
allowed the molecule
to attain its initial distortion in the volume of the experimental 
 unit cell.

 In a first step we stabilized our samples, at the temperatures quoted 
in Table I, with a constant 
volume algorithm for 30000 time steps (each one of 0.01ps.). Later on,
the runs in the (N,P,T) ensemble were performed by lowering and increasing
 the temperature (up to 500K) in steps of 25 or 50K. At each point of the
phase diagram the samples are stabilized for 20000 to 30000 time steps and 
measured in the following 10ps. Near the phase transitions the stabilization 
times were increased several times.
The sample of $\alpha $-S$_8$ crystals consisted of 3*3*2
orthorhombic cells, a sample of 288 molecules. Some points of the phase
diagram were recalculated with a sample of 4*4*2 cells (512 molecules).
The sample of $\beta $-S$_8$ crystals consisted of 4*4*4 cells (384 
molecules) and that of $\gamma $-S$_8$ crystals of 5*3*5 cells (300
molecules).\\ 

{\bf Results:}

{\bf a) }$\alpha ${\bf  -S}$_8${\bf \ crystals:}

$\alpha $-S$_8$ crystallizes, at STP, in the orthorhombic space group F$_d$
(D$_{2h}^{24}$) with 16 molecules in a face-centered unit cell, the 
primitive cell contains 4 molecules at C$_2$ sites. Fig.1a includes part
of our sample at 300K and shows the structure in the [110] direction.
 The molecular rings lie
parallel to the crystallographic $c$ axis and are alternately oriented
parallel to (110) and (1\=10) crystalline planes. Table I shows 
the experimental lattice parameters and our calculated values, the 
largest difference is found in the value of b axis, 2 \% larger than its
experimental value. The molecular array and their corresponding 
orientations follows the 
experimental data. Our calculated configurational energy is -103.5(2)kJ/mol 
at 300K, from which the heat of sublimation can be estimated as 101.0kJ/mol
and compares well with a measured value of 101.8(2)kJ/mol \cite{hsubalfa}. 
Also our calculated pair distribution function g$_2$(r), not
included here, compares 
well with neutron measurements of crystalline $\alpha $-S$_8$ 
at 300K \cite{egelstaff}. 

Fig. 2 shows, for $\alpha -,$ $\beta -$ and 
$\gamma -S_8$, the calculated configurational energy and volume per
 molecule as a function of temperature. Our simulations reproduce
 the experimental data of the three crystals at STP. 
Unfortunately, our $\alpha $-S$_8$ sample of 288 molecules turns 
out to be unstable for T$\leq $200K, and distorts to a monoclinic cell, 
with molecular locations and orientations closely resembling those
of $\alpha $-S$_8$.  This result was later on confirmed with a 
larger sample of 512 molecules (4*4*2 orthorhombic cells). Below 200K this 
large sample does not distort to monoclinic, but nevertheless its 
configurational energy is higher, and its volume per molecule is
 larger, than those calculated
for the sample of 288 molecules. Also, the fluctuations of the energy 
and volume, of the sample of 512 molecules, are larger for
 T$\leq $200 than their value at 200K.

  The coincidence of the calculated distortion at 200K and the 
experimental fact that crystals of pure S$_8$ are
greenish-yellow at 293K and turn white below 193K, made us review 
the experimental data on this phase. Usually the reference quoted for 
lattice parameters as a function of temperature is 
ref.\cite{alfa2}. These values were obtained using X-ray powder 
diffractrometry, on different samples and using only three lines.
Our calculated powder diffraction patterns, not included here,
were similar for the orthorhombic and monoclinic samples and
the scarce experimental data could be not enough to detect 
the changes.
 Nevertheless, there is a careful single crystal determination 
of the structure of $\alpha $-S$_8$ at 300 and 100K \cite{alfa3},
where the crystalline charge distribution is measured. This study 
clearly disregard any possible structural phase transition and shows
that the simple molecule model, in spite of the amount 
of experimental data that it is able to reproduce, should still 
be improved.\\

{\bf a) }$\beta ${\bf  -S}$_8${\bf \ crystals:}

If $\alpha $-S$_8$ is slowly heated, it shows a phase transition to
monoclinic $\beta $-S$_8$ at 369K, which melts at 393K \cite{steudel}.
The structure has been determined from single crystal data \cite{beta2} at
room temperature; this was possible because, after annealing at 383K, the crystals 
may be maintained at 300K for up to a month. The space group is 
C$_{2h}^{5}$  (P2$_1$/c), with six molecules per unit cell.  Four
of the molecules are located at general positions: $_{-}^{+}$(x,y,z; 
x,1/2-y,1/2+z) with x=0.367, y=0.356 and z=0.072. The other
 two molecules ( 5 and 6) are located at inversion centers: 
(0,0,0) and (0,1/2,1/2), achieving this symmetry by orientational 
disorder. Should the disorder be static, the averaged symmetry is over all
molecules at inversion sites, if the disorder is dynamic, the molecules 
achieve its site symmetry on a time average. In both cases the 
disordered array has a residual entropy of k$_B$/3 ln2 per 
molecule \cite{beta2}.
The experimental and lattice parameters of our sample are included 
in Table I, the molecular
 orientations are as measured in ref. \cite{beta2}.

Fig.2 includes the calculated configurational energy and volume
 per molecule, that
 show no discontinuities as a function of temperature. Instead, 
we find that the reorientation of molecules 5 and 6 freezes at 
low temperatures.  This is in accord with heat capacity measurements
that locate an order-disorder transition transition at 198K \cite{beta1}.
Fig. 1b shows our calculated sample at T=150K, the molecules 
5 and 6 are now related by a C2 symmetry axes.

We find that the orientational disorder of molecules at (0,0,0) and 
(1/2,1/2,1/2), at high temperatures, is of the dynamical type.
 Fig. 3 shows the calculated self correlational functions for 
reorientation of the three inertial molecular axes at 370K. The axis
 perpendicular to the molecular plane librates around its equilibrium 
orientation(z axis in Fig.3), whilst the molecule reorientates within 
the plane. The decay time is 
about 10ps. Furthermore, a histogram of molecular
 orientations, averaged over all molecules and all time steps, shows
that the initial correlation has been totally lost after 20000 
equilibration time steps. For T$\leq $150K we measure 
librations of the three axes (with larger amplitudes within the plane), 
but no reorientational diffusion.\\

{\bf a) }$\gamma ${\bf  -S}$_8${\bf \ crystals:}

 $\gamma $-S$_8$ crystals can be grown from solutions \cite{steudel,meyer1}.
The crystal is monoclinic,
 space group C$_{2h}^{4}$ (P2/c), with four 
molecules per unit cell \cite{gam1}. This pseudo-hexagonal 
structure is denser
 than that of $\alpha $-S$_8$, at the same temperature \cite{meyer1,gam1}. 
Our simulations reproduce this structure (Fig. 1c), and Fig. 2 shows 
that at all temperatures this packing has the higher
 density. Our calculated parameters and experimental data, at 300K, are
included in Table I.
It must be noticed that the experimental density,
 measured at 300K, is 2.19gr/cm$^3$,
5.8\% higher than that of $\alpha $-S$_8$  \cite{meyer1,gam1}. 
Nevertheless, in ref.\cite{gam1} the X-ray measurement determines
a structure with volume per molecule higher 
than that of  $\alpha $-S$_8$, implying
a 2.3\% lower density. It could be possible that the temperature of the 
of the X-ray measurement was higher than 300K.\\
No other experimental data are available for further comparisons.
For this sample we find a large discontinuity in lattice parameters
for T$\geq $400K and molecular diffusion for T$\geq $470K.\\

{\bf d) Phase transitions and the liquid phase:}\\
There is experimental evidence that the phase transitions are obtained when
defects are present, if they are not, metastable states can be maintained for 
days \cite{steudel,meyer1,beta2}.
Our calculations do not reproduce the $\alpha $-$\beta $ transition, but this
fact corresponds with the experimental evidence
 that $\beta $-S$_8$ crystals are 
more easily grown from the melt and not by increasing temperature of an 
$\alpha $-S$_8$ sample.

 The same evidence exists for the solid-liquid phase transitions.
In this case, the experimental data show that the disorder is
generated because S$_8$ molecules start to dissociate when
the temperature of the sample is near the melting point.\\
 At even higher temperatures ( 460K) liquid sulfur is known for its
liquid-liquid phase transition, due to changes in the 
molecular composition of the elemental sulfur molecules Sn.\\

That the disorder is esential to promote the solid-liquid phase 
transition can be checked by simulating a liquid sample of 
S$_8$ molecules. These
simulations should be valid for temperatures above and near the
transition, when the fraction of broken molecules is small. A cubic
sample of 216 disordered molecules was studied in the range
350-450K, following the same procedure as the other samples. In
this case the MD algorithm allowed changes in the volume of
the MD box, but not in its shape. Fig. 4
shows the calculated discontinuity in the configurational
energy and volume per molecule at 400K, near that measured for 
$\alpha $ (386K) and $\beta $ (393K) crystals. For
 T$\geq $400K the sample
is liquid, as can be measured by its diffusion constant.
Comparison of Figs. 2 and 4 show that the solid - liquid 
transition is found if the sample is disordered.\\

{\bf  Conclusions:}

In this paper we study the phase diagram of S$_8$ crystals, at 0kbar pressure,
as given by a simple and flexible molecule model.
Our calculations can be also considered a model study of cyclo-chain 
molecules with a torsional potential of a double well type, at difference 
with hydrocarbon chains which show a triple well torsional potential.\\
It has to be emphasized that this phase diagram is extremely complex and 
there are no previous calculations on it. The proposed molecular model 
is extremely simple but, nevertheless, it gives very good account 
of many experimental facts.
The three crystalline phases are stables. The molecular   
model reproduces the
structure, configurational energy and dynamics of $\alpha $-S$_8$, 
$\beta $-S$_8$ and  $\gamma $-S$_8$ crystals for T$\geq $200K. At STP 
the experimental volume per molecule, lattice parameters and 
molecular array of the three phases compare well with the experimental
data, including the smallest volumen per molecule for 
$\gamma $-S$_8$ crystals.\\
   The calculations also reproduce the orientationally disordered phase of
 $\beta $-S$_8$, that we determine as dynamically disordered. At low 
temperatures we calculate a  stable orientationally ordered phase.\\ 
   The solid-liquid phase transition is found near the
experimental value, but only for a disordered sample. The 
$\alpha $- $\beta $-S$_8$ phase transition is not reproduced by 
either of our two samples. Both facts are in accord with the experimental 
evidence that the transitions are promoted by disorder \cite{meyer1}, 
if not, metastable states can be maintained for days.\\

{\bf Acknowledgement:}
The authors thank CONICET for the grant PIP 0859/98.\\

\newpage

{\bf Table I:} Experimental and calculated lattice parameters: a, b, c (\AA )
and monoclinic angle $\beta $ (deg.). Z is the number of molecules 
in the cell, V (\AA $^3$) is the volume per molecule.
 See the text for a discussion about the $\gamma $-S$_8$ data.\\ \\

$
\begin{tabular}{llllllll}
{\bf Phases} &  &{\bf Z} & {\bf a(\AA )} & {\bf b(\AA )} & {\bf c(\AA )}
 &{\bf $\beta $ (deg.)} & {\bf V(\AA $^3$)}
\\ 
&  &  &  &  &  &  &  \\ 
{\bf$\alpha $-S}$_8$ & Ref. \cite{alfa1} & 16 & 10.4646(1) & 12.8660(1) & 
24.4860(3) & - & 206.046(98) \\ 
(T=300 K) & Calc. &  & 10.34(10) & 13.20(3) & 24.32(7) & - & 
207.3(1.0) \\ 
{\bf$\beta $-S}$_8$ & Ref. \cite{meyer1} & 6 & 10.778 & 10.844 & 10.924 & 95.8$%
^{\circ }$ & 211.70 \\ 
(T=370 K) & Calc. &  & 11.01(11) & 10.753(27) & 10.79(10) & 97.48(07)$%
^{\circ }$ & 211.30(74) \\ 
{\bf$\gamma $-S}$_8$ & Ref. \cite{gam1} & 4 & 8.442(30) & 13.025(10) & 9.356(50)
& 124.98(30)$^{\circ }$ & 210.8(1.3) \\ 
(T=300 K) & Calc. &  & 8.03(13) & 13.16(14) & 8.88(11) & 120(3)$^{\circ
}$ & 202.5(2.5)
\end{tabular}
$

\newpage

{\bf Figures:}

Fig. 1: Our calculated crystals: a) orthorhombic $\alpha $-S$_8$ at 
300K, b) monoclinic 
 $\beta $-S$_8$ at 150K, c) pseudo-hexagonal $\gamma $-S$_8$ 
at 300K.

Fig. 2: Calculated configurational energy and volume per molecule  
as a function of temperature for all calculated samples. The lines
are a guide to the eyes.

Fig. 3: $\beta $-S$_8$ crystals at 370K: 
Reorientational self-correlation functions C$_{20}$ for the inertial 
axes of the two molecules at inversion sites.

Fig. 4: Phase diagram of a disordered cubic sample, with the same
scale of Fig. 2. For T$\geq $400K, the sample is liquid. 

\end{document}